# Reliability of Meta-Analysis Research Claims for Gas Stove Cooking−Childhood Respiratory Health Associations

Warren B. Kindzierski[1], S. Stanley Young[2], and John D. Dunn[3]

[1] Independent consultant, St Albert, Alberta, Canada

[2] CGStat, Raleigh, NC, USA

[3] Emergency physician (retired), Brownwood, TX, USA

Correspondence: Warren B. Kindzierski, Independent consultant, St Albert, Alberta, Canada

**Abstract**

Odds ratios or p-values from individual observational studies can be combined to examine a common cause−effect research question in meta-analysis. However, reliability of individual studies used in meta-analysis should not be taken for granted as claimed cause−effect associations may not reproduce. An evaluation was undertaken on meta-analysis of base papers examining gas stove cooking (including nitrogen dioxide, $NO_2$) and childhood asthma and wheeze associations. Numbers of hypotheses tested in 14 of 27 base papers (52%) used in meta-analysis of asthma and wheeze were counted. Test statistics used in the meta-analysis (40 odds ratios with 95% confidence limits) were converted to p-values and presented in p-value plots. The median (interquartile range) of possible numbers of hypotheses tested in the 14 base papers was 15,360 (6,336−49,152). None of the 14 base papers made mention of correcting for multiple testing, nor was any explanation offered if no multiple testing procedure was used. Given large numbers of hypotheses available, statistics drawn from base papers and used for meta-analysis are likely biased. Even so, p-value plots for gas stove−current asthma and gas stove−current wheeze associations show randomness consistent with unproven gas stove harms. The meta-analysis fails to provide reliable evidence for public health policy making on gas stove harms to children in North America. $NO_2$ is not established as a biologically plausible explanation of a causal link with childhood asthma. Biases – multiple testing and p-hacking – cannot be ruled out as explanation for a gas stove−current asthma association claim. Selective reporting is another bias in published literature of gas stove–childhood respiratory health studies.

## 1. Introduction

Population attributable fraction (PAF) is an epidemiologic measure describing the portion of all cases of a disease in a population attributed to a specific exposure (Mansournia & Altman, 2018). Gruenwald et al. (2023) estimated the PAF for current childhood asthma due to natural gas (gas) stove cooking in United States (US) at 12.7% (95% confidence interval 6.3–19.3%). This estimate was based on risk statistics from a Lin et al. (2013a) meta-analysis about indoor nitrogen dioxide ($NO_2$) effects of gas cooking on asthma and wheeze. Cause−effect science claims made by Lin et al. in their meta-analysis were evaluated here for gas cooking−current asthma and gas cooking−current wheeze associations.

### 1.1 Background

Natural gas use – Gas is popular for cooking in over 40 million homes in the US (Lebel et al., 2022a) and 62% of homes in California (Lebel et al., 2022b). Emissions from gas stove cooking include: water vapor, nitrogen gas ($N_2$), oxides of nitrogen ($NO_2$ + nitric oxide), carbon monoxide, carbon dioxide, and trace amounts of methane, ethylene and aldehydes (Logue et al., 2014; Mullen et al., 2016; Poppendieck & Gong, 2018; Lebel et al., 2020, 2022a,b). Lebel et al. (2020, 2022a) estimate that 75% or more of total methane emissions from gas stoves and storage water heaters may occur when they are off.

Gas consumption from residential, commercial, and industrial use in the US increased by 17% (58.3 to 68.4 Billion cubic feet per day) over the period 2003−2013 (British Petroleum, 2022). During this time, hospitalizations of children aged 0−17 years for asthma decreased by 50% and missed school days of children aged 5−17 years with asthma decreased by 20% in the US (CDC, 2018).

Recent gas stove health policy and proposed regulations – Krasner et al. (2021) suggest warning labels are needed for gas cooking stoves to address home indoor air quality. Position statements of the American Medical Association (2022) and the American Public Health Association (2022) take a precautionary approach recommending replacement of gas stoves with electric stoves in homes.

In an early February 2023 notice of proposed rulemaking, the U.S. Department of Energy (DOE) proposed new energy conservation standards for stove appliances. These conservation standards limit how much energy gas and electric stove tops may consume in a year. For gas cooking tops, the conservation standards limit maximum allowable integrated annual energy consumption to 1,204 kBtu (kilo-British thermal units) per year; referred to as the "EL 2" standard (DOE, 2023a).

DOE (2023b) later released an updated analysis which suggests that half of the current gas cooking tops on the market would be prohibited under the standard. Environmental advocacy groups Appliance Standards Awareness Project and American Council for an Energy-Efficient Economy predict the rulemaking may lead to ~30% less energy consumption compared to least-efficient gas stoves in use today (deLaksi, 2023).

*1.2 Gas Stove ($NO_2$)−Asthma Connection*

Asthma is a complex, chronic bronchial inflammatory phenomenon attributed to an allergic reaction characterized by airway spasm and swelling of airway walls along with hyper responsiveness (Noutsios & Floros, 2014). Allergens (molecules that stimulate allergic reactions) are the basic issue in asthma morbidity (Froidure et al., 2016). Pollen and plant parts; biological fragments shed from furry animals, rodents, cockroaches, dust mites; and fungal detritus can cause allergenic asthma reactions (Kanchongkittiphon et al., 2015; Kader et al., 2018). This is because the airways have an allergic response to carbon-based molecules.

$NO_2$ molecules are not carbonaceous and cannot trigger allergic reactions because they are not allergens. Individuals can also have allergies to ingested or contact substances that can produce asthma attacks. Classic examples are bee sting venom or peanut allergy that can produce asthma attacks and even the more severe allergic reaction – anaphylaxis.

Reactive airways disease as a phenomenon is a spastic and congested airway condition that is brought on by airway hypersensitivity. This would include asthma. But it might be caused by asthma or inflammation of the airways separate from allergic asthma (e.g., continuous exposure to inhaled tobacco smoke or chemically irritating and damaging industrial gases). The difference between reactive airways disease and asthma is demonstrated clinically by history and allergy testing (AAFA, 2021). Asthma is also demonstrated by cold air, exercise, and methacholine challenge testing (AAFA, 2021; Mayo Clinic, 2022).

Controlled exposure (chamber) studies of humans inhaling $NO_2$ report airway responsiveness in asthmatics (e.g., Ezratty et al., 2014; Brown, 2015). Several published reviews of chambers studies note ambiguity (uncertainty) in the role of $NO_2$ on airway responsiveness in asthmatics (e.g., Goodman et al., 2009, 2017, Hesterberg et al., 2009). Given the nature of asthmatics and their airway hyperactivity and sensitivity, chamber studies do not establish $NO_2$ exposures as proof of causation of asthma.

A recent, official American Thoracic Society (ATS) report reaffirms the ambiguous role of $NO_2$ on childhood asthma. The report states (Thurston et al., 2020): "*It is unclear whether direct effects of $NO_2$ ... explain the causal link with asthma*". The ATS report recognizes epidemiolocal associations of $NO_2$ with childhood asthma. However, it is silent on biological plausibility of a $NO_2$−childhood asthma link. There is no dispute about the role of allergy in causing asthma.

*1.3 Meta-Analysis*

Meta-analysis is a procedure for statistically combining data (test statistics) from individual studies that address a common research question or claim (Egger et al., 2001). An example of a common question or claim (i.e., cause−effect science claim) in a meta-analysis is whether a risk factor or intervention is causal of a disease.

A meta-analysis evaluates a claim by taking a test statistic (e.g., risk ratio, odds ratio, hazard ratio, etc.) along with a measure of its reliability (e.g., confidence interval) from multiple individual studies (base papers) from literature. The test statistics are combined to give, theoretically, a more reliable estimate of cause−effect (Young & Kindzierski, 2019). It initially involves a systematic review of literature using specific methods to identify, select, and critically appraise relevant research, and to compile and analyze data from the selected studies (Moher et al., 2009). The meta-analysis component then selects and combines test statistics of the identified studies.

One aspect of replication – performance of another study statistically confirming the same hypothesis or claim – is a foundation of science and replication of research is essential before cause−effect claims can be asserted (Moonesinghe et al., 2007). A well-designed and conducted meta-analysis is highly ranked in the medical evidence-based pyramid – similar to well-designed and conducted randomized trials, and above observational (case–control and cohort) studies (van Wely, 2014; Murad et al., 2016; Herner, 2019).

A hypothesis of meta-analysis is that test statistics taken from base papers for analysis are unbiased estimates of cause−effect (Boos & Stefanski, 2013). Given these characteristics, independent evaluation of published meta-analysis on a common research question has been used to assess the statistical reproducibility of a claim coming from that field of research (Young & Kindzierski, 2019; Kindzierski et al., 2021; Young et al., 2022a; Young & Kindzierski, 2022, Young & Kindzierski, 2023).

*1.4 Irreproducible Research*

Researchers have increasingly recognized that far too much published research is irreproducible or false (e.g., Ioannidis, 2005; Ioannidis et al., 2011; Young & Karr, 2011; Begley & Ellis, 2012; Keown, 2012; Begley & Ioannidis, 2015; Iqbal et al., 2016; Randall & Welser, 2018; Stodden et al., 2018). Irreproducible research occurs in a wide range of scientific disciplines – general medicine, clinical sciences, oncology, nutrition, biology, psychological sciences (Young & Kindzierski, 2022).

NASEM (2016) cites a number of factors that contribute to irreproducible research. These include:

- Insufficient training in experimental design.
- Misaligned incentives for publication and the implications for university tenure.
- Intentional manipulation.
- Poor data management and analysis.
- Inadequate instances of statistical inference.

The irreproducibility problem partly originates from scientists exploring large numbers of hypotheses and using multiple models without statistical correction in their analysis. This is referred to as multiple testing, multiple testing and multiple modelling, or multiplicity (Westfall & Young, 1993; Sainani, 2009; Patel & Ioannidis, 2014; Young & Kindzierski, 2019).

Multiple testing can occur when scientists use their data set to test multiple outcomes, multiple predictors, different population subgroups, multiple statistical cause−effect models, or multiple confounders to cause−effect associations. It increases the likelihood of making a type I (false positive) error.

*1.5 Study Objectives*

Large observational data sets are potentially available to epidemiology researchers investigating cause−effect associations in humans. These investigations require strong statistical evidence to establish useful, replicable, and understandable cause−effect associations and research claims made from these associations. Strong statistical evidence is robust relative risks, RRs (or odds ratios, ORs), and confidence intervals that do not include 1.0. RRs (or ORs) greater than 2 to 3 are usually recommended as robust enough in observational (uncontrolled) population studies to rule out bias and confounding (Doll & Peto, 1981; Ahlbom et al., 1990; Taubes, 1995; Bonita et al., 2006; Federal Judicial Center, 2011).

Epidemiological studies discussed here are not founded on proven biological plausibility of $NO_2$ causing childhood asthma. They are founded on a concept of what may be a cause of childhood respiratory disease and harm, e.g., gas stoves and whatever emissions they create. Accepting this as a concept, meta-analysis was used by Lin et al. (2013a) to claim the presence of an evidentiary correlation between $NO_2$ from gas stove cooking and childhood asthma.

Our evaluation involved using search space analysis (Young & Kindzierski, 2019) and p-value plotting (Schweder & Spjøtvoll, 1982) to assess the reproducibility of two Lin et al. meta-analysis cause−effect research claims. These claims were for gas stove cooking−current asthma and gas stove cooking−current wheeze associations. This was done to judge reliability of the Gruenwald et al. (2023) PAF estimate for current childhood asthma in the US due to gas stove cooking.

## 2. Methods

*2.1 Data Sets*

Lin et al. (2013a) undertook a meta-analysis to quantify the association of indoor $NO_2$ and gas stove cooking with childhood asthma and wheeze. They ran computer searches in two online databases (PubMed and the ISI Web of Knowledge) for the period 1977 up to 31 March 2013. They identified 1,064 articles from the databases, from which 329 duplicates and/or non-English articles were excluded. They then selected 735 articles for abstract review and from this list, fully reviewed 171 articles.

Lin et al. ultimately selected 41 articles (base papers) using asthma and wheeze as health outcomes for data abstraction, quality assessment and meta-analysis. The meta-analysis was published on 20 August 2013 in the International Journal of Epidemiology. As of 1 February 2023, the meta-analysis was listed as the second 'most read' article on their website (https://academic.oup.com/ije).

Using test statistics drawn from the 41 base papers, Lin et al. initially examined eight different cause−effect associations (outcomes) in their meta-analysis. These included:

1) Gas cooking−current asthma.
2) Gas cooking−lifetime asthma.
3) Indoor $NO_2$−current asthma.
4) Indoor $NO_2$−lifetime asthma.
5) Gas cooking−current wheeze.
6) Gas cooking−lifetime wheeze.
7) Indoor $NO_2$−current wheeze.
8) Indoor $NO_2$−lifetime wheeze.

Our evaluation focused on two of eight outcomes: gas cooking−current asthma and gas cooking−current wheeze. Both outcomes had large enough data sets to allow informative p-value plots. In addition, gas cooking−current asthma was of interest because the Gruenwald et al. (2023) PAF is for current childhood asthma from gas stove cooking. Gas cooking−current wheeze was of interest as it provides an effective control check for p-value plotting.

Lin et al. used the DerSimonian & Laird (1986) random effects method for combining test statistics in their meta-analysis. They combined test statistics from 11 base papers for a gas cooking−current asthma association and declared the result significant (OR = 1.42; 95% confidence interval, CI = 1.23–1.64). Also, they combined test statistics from 22 base papers on a gas cooking−current wheeze association and declared the result non-significant (OR 1.07, 95% CI 0.99–1.15).

Secondly, Lin et al. conducted additional statistical analysis (re-analysis) on a subset of their data using the same methods. This re-analysis was on six of the eight outcomes (gas cooking−lifetime wheeze and indoor $NO_2$−lifetime wheeze associations were excluded). Also, this analysis excluded data for two base papers performed on asthmatics only and it included data for studies with confounder adjustments.

Thirdly, Lin et al. conducted more statistical analysis on a different subset of their data with the same methods, this time focusing on just three of the eight outcomes: gas cooking−current asthma, gas cooking−lifetime asthma, and gas cooking−all asthma. This re-analysis examined differences in four factors: age groups (3 categories), study region (3 categories), proportion of gas cooking (2 categories), and year of publication (2 categories). As shown later, subgroup analyses are problematic for causal claims.

Gruenwald et al. (2023) selectively used test statistics from Lin et al. re-analyses of study region differences. They combined two (2) test statistics – one representing North American region studies (OR 1.36, 95% CI 0.76–2.43) and one representing European region studies (OR 1.34, 95% CI 1.13–1.60) – using inverse variance weighting. They excluded Asia-Pacific region study test statistics. The combined estimate using the two test statistics was OR 1.34, 95% CI 1.12–1.57.

*2.2 Search Space Analysis (Numbers of Statistical Hypotheses Tests) of Base Papers*

Numbers of statistical hypotheses tested in a study needs to be estimated (counted) to assess the potential for multiple testing bias (Makin & de Xivry, 2019). Lin et al. (2013a) used 11 base papers in their meta-analysis of current asthma and 22 base papers in their meta-analysis of current wheeze. Attempts were made to obtain digital copies of all 11 base papers related to current asthma to count the number of hypotheses tested in each paper.

Only ten of 11 papers could be obtained for counting. Six of these were also used for the current wheeze meta-analysis. An additional four base papers used for current wheeze – two published before 2000 and two after 2000 – were further randomly selected for counting. Appendix 1 lists all 14 Lin et al. base papers used for counting.

The search space – number of hypotheses, $N_H$, tested – in base papers used by Lin et al. was estimated (counted) as follows. Observational studies mostly use a direct statistical analysis strategy on data collected – e.g., what outcomes are related to predictors or risk factors. If a data set contains "O" outcomes and "P" predictors, O × P possible hypotheses can be tested. A covariate "C" (also called an adjustment factor) – such as age, weight, height, sex, etc. – can be included as a yes/no adjustment to see how it may modify each of the O × P hypotheses. Here a covariate is included or excluded; and a multiplier of 2 is assumed for each covariate considered.

The 14 base papers were carefully read and counted for outcomes (O), predictors (P), and covariates (C). The search space (number of hypotheses considered in a base paper, $N_H$) can be approximated as $O \times P \times 2^C$. Each base paper was then re-read with interest whether a paper:

- Mentioned multiple testing in different forms (i.e., multiple hypotheses, multiple testing, multiple comparisons, multiplicity).
- Mentioned correcting for this bias.

Figure 1 provides an example of how the search space, $N_H$, is estimated in an observational study after it is carefully read, and outcomes, predictors, and covariates are identified. The counts in Figure 1 and for the Lin et al. base papers are lower bound approximations (Young & Kindzierski, 2019), as they are only based on information that is reported in each study.

A conventional threshold for statistical significance in most science disciplines is a p-value < 0.05. A false positive result should occur 5% of the time by chance using this threshold when multiple testing is undertaken on the same data set in the absence of any statistical corrections (Young & Kindzierski, 2022; Young et al., 2022b). The expected number of false positive (chance) findings in the Moshammer et al. (2010) study is estimated as $0.05 \times N_H = 0.5 \times 461{,}440 = 23{,}072$.

---

Moshammer et al. (2010) gas stove cooking–lung function study.

*Background* – Meta-analysis of pooled data for 24,000 children aged 6–12 years from nine western countries (United States, Canada, Austria, Germany, the Netherlands, Poland, Hungary, Czech Republic, Slovakia).

*Outcomes (O)*

Seven lung function measures – forced vital capacity (FVC), forced expiratory volume in 1 s ($FEV_1$), peak expiratory flow (PEF), maximum mid-expiratory flow (MMEF), maximal expiratory flow (MEF) at 25% FVC, MEF at 50% FVC, and MEF at 75% FVC. All measures obtained using standard lung function test protocols of the American Thoracic Society (North America) or the European Respiratory Society (Europe).

Four statistical model formulations for each lung function measure – three 'basic' statistical models (using a data set for all 9 countries, using a data set for all countries excluding Austria, using a data set for all countries excluding United States), and an 'adjusted' statistical model using a data set for all countries.

∴ O = 7 lung function measures × 4 statistical models = 28.

*Predictors (P)*

Stove type (either gas or electric); ∴ P = 1.

*Covariates (C)*

'Basic' statistical models corrected for 7 confounders – age, weight, height, sex, seasonal trend, technician and/or instrument (if the study center used more than one) and study area.

'Adjusted' statistical model corrected for 16 confounders – 7 previous confounders + 9 additional: smoking in pregnancy, recent respiratory infections, current medication, maximal parental education, household crowding, unventilated gas/oil/kerosene heater, mold, birth order, and 'ever had a pet'.

∴ C = 7 (basic statistical models) and 16 (adjusted statistical model).

*Search space ($N_H$)*

Basic statistical models, sub-total $N_H = O \times P \times 2^C = 7(3) \times 1 \times 2^7 = 2{,}688$.
Adjusted statistical model, sub-total $N_H = O \times P \times 2^C = 7(1) \times 1 \times 2^{16} = 458{,}752$.
Total $N_H = 2{,}688 + 458{,}752 = 461{,}440$.

Figure 1. Example of counting numbers of hypotheses tests in gas stove cooking observational studies

*2.3 Numbers of Statistical Hypotheses Tests in Cohort Population Data Sets*

Two base papers used in the current asthma meta-analysis (Willers et al., 2006; Lin et al., 2013b) utilized the PIAMA birth cohort data set for their studies. PIAMA (Prevention and Incidence of Asthma and Mite Allergy) is a prospective birth cohort started in 1998 with 3,963 newborns in the Netherlands (Brunekreef et al., 2002).

A compounding multiple testing problem can occur with repeated use of cohort population data sets (Young & Kindzierski, 2022; Young et al., 2022b). It can be a lengthy and costly process to establish and follow a new cohort. However, it can be more efficient and less costly to use an existing cohort and simply add new measurements/observations and research questions (hypotheses) to it. It is possible over time to have many hypotheses tested on a given cohort as data for the cohort can be used repeatedly by independent researchers.

A single published study for a particular cohort data set may only address the tip of the problem of numbers of hypotheses tested. Overall, there may be many other hypotheses at issue considering that the same cohort data set can be used many times for research. Many publications in literature for a single cohort data set suggests large number of hypotheses examined with possibilities of large numbers of false positive (chance) findings present overall in literature (Young & Kindzierski, 2022; Young et al., 2022b).

The potential multiple testing problem with cohort data sets was explored with the PIAMA birth cohort. The Advanced Search Builder capabilities of the PubMed search engine was used to identify the number of studies that used the PIAMA birth cohort data set and were published since its inception ($N_C$). The term PIAMA[Title/Abstract] was used for the period 2000-2023 (search performed 1 February 2023). The total number of hypotheses tested on a cohort data set can be estimated as $N_C \times N_{H,median}$, where $N_{H,median}$ is the median (or typical) number of hypotheses tested in a published cohort study.

*2.4 P-Value Plots*

In epidemiology it is traditional to use RRs (or ORs) and CIs instead of p-values from a hypothesis test to demonstrate or interpret statistical significance. RRs (or ORs) and CIs and p-values are constructed from the same data set, and they are interchangeable. Altman and Bland (2011a,b) show how one can be computed from the other. Alternatively, commercial statistical software packages – e.g., SAS or JMP (SAS Institute, Cary, NC), STATA (StataCorp LLC, College Station, TX) – can be used. Here, p-values were computed using JMP statistical software for all Lin et al. (2013a) data used in their meta-analyses of current asthma and current wheeze.

P-value plots (Schweder & Spjøtvoll, 1982) were constructed to examine the distribution of the set of p-values for the studies. The p-value is a random variable derived from a distribution of the test statistic used to analyze data and to test a null hypothesis (Hung et al., 1997). In well-designed and conducted studies, the p-value is distributed uniformly over the interval 0 to 1 under the null hypothesis, no effect, regardless of sample size (Schweder & Spjøtvoll, 1982).

Properly scaled, a distribution of p-values plotted against their ranks in a p-value plot should form a 45-degree line when there are no effects (Schweder & Spjøtvoll, 1982; Hung et al., 1997; Bordewijk et al., 2020). Researchers can use a p-value plot to visually examine the heterogeneity of the test statistics combined in a meta-analysis (Young & Kindzierski, 2019, 2022, 2023).

The p-value plots constructed here were interpreted as follows (Young & Kindzierski, 2023):

- Computed p-values are ordered from smallest to largest and plotted against the integers, 1, 2, 3,…
- If p-value points follow an approximate 45-degree line, it is concluded that test statistics result from a random (chance) process and the data support the null hypothesis of no significant association or effect.
- If p-value points follow approximately follow a line with a flat/shallow slope, where most (the majority) of p-values are small (< 0.05), then the data provides evidence for a real, statistically significant, association or effect.
- If p-value points exhibit a bilinear shape (divides into two lines) and bias is present or can be established, the data is consistent with a mixture and a general (overall) research claim is not supported. Further, a small p-value reported for an overall claim in the meta-analysis may not be valid (Schweder & Spjøtvoll, 1982).

Examples of p-value plots are provided in Appendix 2 to assist in interpretation of p-value plots constructed here. P-value plots in Appendix 2 represent 'plausible null' and 'plausible true alternative' hypothesis outcomes based on meta-analyses of observational and/or randomized trial data sets. Meta-analyses p-values supporting a plausible null hypothesis plot as an approximate 45-degree line. Whereas meta-analyses p-values supporting a plausible true alternative hypothesis plot as a line with a flat/shallow slope, where most (the majority) of p-values are small (< 0.05).

The distribution of the p-value under the alternative hypothesis – where p-values are a measure of evidence against the null hypothesis – is a function of both sample size and the true value or range of true values of the tested parameter (Hung et al., 1997). P-value plots in Appendix 2 show distinct (single) sample distributions for each condition – i.e., null (chance or random) associations and true effects between two variables tested. P-value plots displaying patterns outside of those shown in Appendix 2 should be treated as ambiguous (uncertain). A research claim based on ambiguous evidence is unproven.

## 3. Results

### 3.1 Numbers of Hypotheses Tested in Base Papers (Counting)

A total of 14 base papers were counted to estimate numbers of hypotheses tested. This, effectively, resulted in 10 of 11 base papers (91%) for current asthma and 10 of 22 base papers (45%) for current wheeze being counted. Overall, 14 (or 52%) of 27 individual base papers used by Lin et al. (2013a) were counted for the current asthma and current wheeze meta-analyses.

A 5 to 20% sample from a population whose characteristics are known is considered acceptable for most research purposes as it provides an ability to generalize for the population (Creswell 2003). It is reasonable to accept the Lin et al. judgment that their systematic review process selected 41 base papers with sufficiently consistent (i.e., known) characteristics for meta-analysis. On this basis, counting numbers of hypotheses tested in 52% of base papers used in their current asthma and current wheeze meta-analyses is considered more than adequate to assess the potential for multiple testing bias.

Statistics for possible numbers of hypotheses tested in the 14 base papers are presented in Table 1. The median number (interquartile range, IQR) of possible numbers of hypotheses tested (Search Space) of the 14 base papers was 15,360 (IQR 6,336−49,152). Given the large numbers of hypotheses tested, statistics drawn from the base papers and used for meta-analysis by Lin et al. are likely biased.

For comparison purposes, possible numbers of hypotheses tested in environmental epidemiology health effects studies of $NO_2$ and other air quality parameters have been reported elsewhere:

- Asthma exacerbation, 17 base papers: median (IQR) = 15,360 (1,536–40,960), (Kindzierski et al., 2021).
- Development of asthma, 19 base papers: median (IQR) = 13,824 (1,536−221,184), (Young et al., 2022a).
- Heart attack, 34 base papers: median (IQR) = 12,288 (2,496−58,368) (Young & Kindzierski, 2019).

Given these numbers, it appears routine for researchers to be able to test for many statistical hypotheses in environmental epidemiology health effects studies of $NO_2$. Five percent of 15,360 possible hypotheses tested in a typical base paper used by Lin et al. (i.e., the median $N_H$ in Table 1) equals 768 chance findings that may be mistaken for real (statistically significant) results. Also, none of the 14 base papers reviewed (0%) made mention of correction for multiple testing nor did they provide any explanation if no multiple testing procedure was used.

### 3.2 Number of Hypotheses Tested in Cohort Studies

Regarding the two Lin et al. (2013a) base papers using the same PIAMA birth cohort data set for their studies, the PubMed search identified 146 listings (publications) for this cohort data set since 2000. Abstract for these listings were read online. Of these 146 listings, 107 explicitly used the PIAMA birth cohort data set to investigate various risk factor research hypotheses (= $N_C$).

The Young at al. (2022a) study mentioned above examined effects of $NO_2$ and other air quality parameters on development of asthma using cohort population data sets. Specifically, Young et al. counted possible numbers of hypotheses tested in 19 base papers representing 13 different population cohorts, including the PIAMA birth cohort. The median count reported by Young et al. (i.e., $N_{H,median}$ = 13,824) was used to estimate possible numbers of false positive findings in published PIAMA birth cohort studies. This was estimated as 5% × $N_C$ × $N_{H,median}$ = 5% × 107 × 13,824 = 73,958 (almost 74,000). This represents the number of false positive (chance) findings that may be mistaken for real results across 107 PIAMA cohort studies in published literature.

Bolland and Grey (2014) cautioned researchers about multiple testing limitations of cohort studies. The example they used was on published research on the Nurses' Health Study cohort (NHS, Colditz & Hankinson, 2005).

Table 1. Characteristics of 14 base papers from Lin et al. (2013a) meta-analysis

| 1st Author Year | Region | Outcomes | Predictors | Covariates | Search Space, $N_H$ |
|---|---|---|---|---|---|
| Current asthma: | | | | | |
| Carlsten 2011 | North America | 3 | 1 | 13 | 24,576 |
| Diette 2007 | North America | 1 | 10 | 5 | 320 |
| Hessel 2001 | North America | 3 | 9 | 8 | 6,912 |
| Tavernier 2005 | England | 1 | 7 | 13 | 57,344 |
| Current asthma + wheeze: | | | | | |
| Behrens 2005 | Europe | 16 | 7 | 5 | 3,584 |
| Dekker 1991 | North America | 6 | 4 | 9 | 12,288 |
| Lin 2013b | Europe | 198 | 3 | 9 | 304,128 |
| Melia 1977 | Europe | 12 | 1 | 9 | 6,144 |
| Spengler 2004 | Russia | 20 | 10 | 9 | 102,400 |
| Willers 2006 | Europe | 3 | 3 | 11 | 18,432 |
| Current wheeze: | | | | | |
| Belanger 2006 | North America | 8 | 1 | 10 | 8,192 |
| Burr 1999 | Europe | 10 | 1 | 9 | 5,120 |
| Strachan 1996 | Europe | 9 | 1 | 11 | 18,432 |
| Wong 2004 | Hong Kong, mainland China | 2 | 2 | 15 | 131,072 |
| | minimum= | 1 | 1 | 5 | 320 |
| | lower quartile= | 3 | 3 | 8 | 6,336 |
| | median= | 5 | 6 | 9 | 15,360 |
| | upper quartile= | 15 | 9 | 11 | 49,152 |
| | maximum= | 198 | 10 | 13 | 304,128 |
| | mean= | 26 | 6 | 9 | 49,925 |
| | n= | 14 | | | |

Notes: Search Space, $N_H$ = Outcomes × Predictors × $2^{Covariates}$.

Bolland and Grey stated: "*Investigators have published more than 1000 articles on the NHS, at a rate of more than 50 papers/year for the last 10 years. ... To date, more than 2000 hypotheses have been tested in these papers, and it seems likely that the number of statistical tests carried out would be in the tens of thousands. ... Given the volume of hypotheses assessed and statistical tests undertaken, it seems likely that many results reported in NHS publications are false positives, and that the use of a threshold of p=0.05 for statistical significance is inappropriate without consideration of multiple statistical testing. ... We suggest that authors of observational studies should report how many hypotheses have been tested previously in their cohort study, together with an estimate of the total number of statistical tests undertaken.*"

*3.3 P-value Plots*

Tables 2 and 3 show ORs, CIs, and p-values estimated for Lin et al. (2013a) meta-analysis of gas stove−current asthma and gas stove−current wheeze. Altogether, 40 ORs and CIs (13 for current asthma, 27 for current wheeze) were converted to p-values and presented in p-value plots. Figures 2 and 3 show p-value plots for Lin et al. meta-analysis of gas stove−current asthma and gas stove−current wheeze.

Figures 2 and 3 show no evidence of distinct (single) sample distributions for true effects between two variables (i.e., p-value points forming a line with a flat/shallow slope, where the majority of p-values are small, < 0.05). Appendix 2 shows examples of p-value plots with true effects between two variables in observational studies. Both plots (Figures 2 and 3) show evidence of distinct sample distributions for null effects between two variables. These plots are consistent with chance or random associations and unproven harms for gas stove cooking.

Table 2. Gas stove−current asthma effect odds ratios (ORs), confidence intervals (CIs) and p-values estimated for

Lin et al. (2013a) meta-analysis

| # | Base paper 1st author, year | OR | 5% CI | 95% CI | p-value |
|---|---|---|---|---|---|
| 1 | Melia, boys, 1977 | 1.48 | 0.9 | 2.43 | 0.2188 |
| 2 | Melia, girls, 1977 | 1.53 | 0.79 | 2.96 | 0.3384 |
| 3 | Dekker, 1991 | 1.95 | 1.14 | 2.68 | ***0.0156*** |
| 4 | Hessel, 2001 | 1.7 | 1 | 3.1 | 0.1913 |
| 5 | McConnell, no wheeze, 2002 | 1.3 | 0.8 | 1.9 | 0.2850 |
| 6 | McConnell, wheeze, 2002 | 1.2 | 0.7 | 2 | 0.5465 |
| 7 | Spengler, 2004 | 2.28 | 1.04 | 5.01 | 0.2063 |
| 8 | Behrens, 2005 | 0.77 | 0.17 | 3.46 | 0.7841 |
| 9 | Tavernier, 2006 | 0.69 | 0.24 | 1.95 | 0.4773 |
| 10 | Willers, 2006 | 1.5 | 0.9 | 2.49 | 0.2177 |
| 11 | Diette, 2007 | 0.84 | 0.47 | 1.48 | 0.5346 |
| 12 | Carlsten, 2011 | 1.4 | 0.6 | 3.6 | 0.6012 |
| 13 | Lin, 2013 | 1.29 | 0.98 | 1.69 | 0.1093 |

Notes: 5%/95% CI = 5th/95th percentile confidence intervals; bolded p-value <0.05.

Table 3. Gas stove−current wheeze effect odds ratios (ORs), confidence intervals (CIs) and p-values estimated for Lin et al. (2013a) meta-analysis

| # | Base paper 1st author, year | OR | 5% CI | 95% CI | p-value |
|---|---|---|---|---|---|
| 1 | Melia, boys, 1977 | 1.11 | 0.87 | 1.4 | 0.4159 |
| 2 | Melia, girls, 1977 | 1.55 | 1.16 | 2.07 | ***0.0178*** |
| 3 | Ware, 1984 | 1.1 | 0.97 | 1.24 | 0.1465 |
| 4 | Hosein, boys, 1989 | 0.61 | 0.4 | 0.94 | ***0.0046*** |
| 5 | Hosein, girls, 1989 | 0.64 | 0.38 | 1.08 | ***0.0438*** |
| 6 | Dekker, 1991 | 1.04 | 0.77 | 1.42 | 0.8094 |
| 7 | Strachan, 1995 | 0.86 | 0.61 | 1.23 | 0.3761 |
| 8 | Volkmer, 1995 | 1.16 | 1.01 | 1.32 | ***0.0430*** |
| 9 | Butland, 1997 | 1.34 | 0.95 | 1.89 | 0.1562 |
| 10 | Maier, 1997 | 0.9 | 0.5 | 1.6 | 0.7216 |
| 11 | Garrett, 1998 | 1.79 | 0.8 | 3.99 | 0.3301 |
| 12 | Burr, 1999 | 1.03 | 0.97 | 1.1 | 0.3657 |
| 13 | Zacharasiewicz, 1999 | 1.16 | 0.92 | 1.46 | 0.2454 |
| 14 | Pikhart, 2000 | 0.87 | 0.65 | 1.17 | 0.3271 |
| 15 | Ponsonby, 2001 | 1.08 | 0.75 | 1.55 | 0.6951 |
| 16 | Belanger, asthmatic mother, 2003 | 0.98 | 0.57 | 1.66 | 0.9427 |
| 17 | Belanger, non-asthmatic mother, 2003 | 1.31 | 0.91 | 1.88 | 0.2103 |
| 18 | Spengler, 2004 | 1.06 | 0.86 | 1.31 | 0.6012 |
| 19 | Wong, 2004 | 1.4 | 0.85 | 2.32 | 0.2828 |
| 20 | Behrens, boys, 2005 | 0.55 | 0.31 | 0.98 | ***0.0085*** |
| 21 | Behrens, girls, 2005 | 1.52 | 0.93 | 2.47 | 0.1856 |
| 22 | Belanger, multifamily home, 2006 | 2.27 | 1.15 | 4.47 | 0.1337 |
| 23 | Belanger, single-family home, 2006 | 0.61 | 0.35 | 1.05 | ***0.0290*** |
| 24 | Willers, 2006 | 0.99 | 0.74 | 1.32 | 0.9461 |
| 25 | Wong, 2007 | 1.68 | 1.03 | 2.75 | 0.1212 |
| 26 | Mitchell, 2009 | 0.93 | 0.81 | 1.07 | 0.2912 |
| 27 | Lin, 2013 | 1.06 | 0.92 | 1.22 | 0.4330 |

Notes: 5%/95% CI = 5th/95th percentile confidence intervals; bolded p-value <0.05.

When observational data are used in cause−effect health studies, strong statistical evidence is required to establish informative associations. This is also the case for research claims made from these associations. For a claim based on observational data to be considered true, it must overcome randomness (i.e., a statistical outcome due to chance). Both p-value plots show sample distributions consistent with randomness, not true effects.

Keep in mind that sample statistics (and the respective base papers) used in meta-analysis were identified by Lin et al. after rigorous systematic review (screening) of the scientific literature. That is to say, they were judged by Lin et al. as the most suitable and appropriate test statistics (and base papers) to address the research question of a gas stove cooking−current asthma (current wheeze) association.

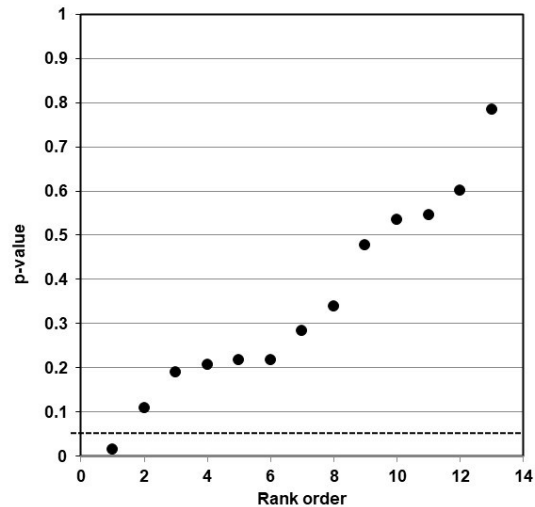

Figure 2. P-value plot for Lin et al. (2013a) meta-analysis of gas stove−current asthma effect
Notes: 13 p-values overall; 1 p-value is < 0.05; 12 p-values are nulls (> 0.05).

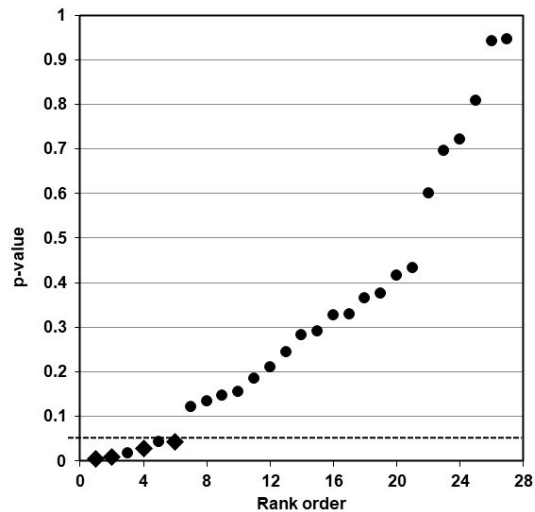

Figure 3. P-value plot for Lin et al. (2013a) meta-analysis of gas stove−current wheeze effect
Notes: 27 p-values overall; 6 of 27 p-values are significant (< 0.05); 4 p-values < 0.05, diamonds, are for negative effects – gas cooking decreases childhood wheeze; 21 of 27 p-values are nulls (> 0.05).

## 4. Discussion

*4.1 Weak Statistical Associations*

Almost 20 years ago, Weed (2006) stated in the International Journal of Epidemiology: "*epidemiology was once a legitimate science of disease causation, but no longer*" and "*we have found all the strong associations, with only the weak left to be discovered*". Notwithstanding this, influences of environmental factors to disease and death continue to be of interest to epidemiologists. None of the meta-analytic test statistics computed by Lin et al. (2013a) are robust enough to rule out bias and confounding (i.e., ORs greater than 2 to 3).

Further, test statistics used for making cause−effect claims must overcome randomness. Test statistics used by Lin et al. do not overcome randomness when presented in p-value plots. Rather, they support unproven harms for $NO_2$ from gas stove cooking and current asthma (and current wheeze).

In 2000, the Committee on the Assessment of Asthma and Indoor Air of the Institute of Medicine (IOM, 2000) evaluated and summarized scientific evidence for interactions between indoor $NO_2$ and development and/or exacerbation of asthma. At the time IOM concluded: "*There is sufficient evidence of an association between brief high-level exposures to $NO_2$ and increased airway responses to both nonspecific chemical irritants and inhaled allergens among asthmatic subjects. These effects have been observed in human chamber studies at concentrations that may occur only in poorly ventilated kitchens with gas appliances in use.*"

Since 2000, additional research on indoor exposures and asthma had been completed. Kanchongkittiphon et al. (2015) published an update of the IOM (2000) assessment. Kanchongkittiphon et al. stated: "*Findings on gas stove use and exacerbation of asthma are too inconsistent to demonstrate associations*" and "*There is limited or suggestive evidence of an association between $NO_2$ and exacerbation of asthma, although this association may be attributable to confounding by other consistently correlated emissions from gas stoves*," and "*There is inadequate or insufficient evidence to determine whether an association exists between gas stove use and exacerbation of asthma.*"

Raju et al. (2020) and Garcia et al. (2021) published reviews on indoor air quality−childhood respiratory health studies. These reviews are recent with relevant features to our study – i) a focus on childhood asthma in developed countries (North America and Western Europe) and, ii) highlight gas stove cooking among sources.

Similar to Kanchongkittiphon et al. (2015), Garcia et al. (2021) noted that controlled human $NO_2$ exposure studies and individual observational studies of inferred $NO_2$ exposures show insufficient evidence. This is due to inconsistent findings (i.e., studies showing positive and negative (null) associations) for gas stove cooking−asthma outcomes, particularly lung function in children. Pulmonary (lung) function examination is the main test physicians use to confirm asthma in children 5 years or older (Rothe et al., 2018; AAFP, 2020).

An example of inconsistent findings mentioned by Garcia et al. is the Moshammer et al. (2010) gas stove cooking–lung function study shown in Figure 1. The Moshammer et al. study exhibits evidence of multiple testing bias ($N_H$ = 461,440). Of the seven outcomes examined by Moshammer et al., only two – FVC and $FEV_1$ – showed small gas stove cooking−childhood lung function effects (<1% reduction using electric versus gas stove). These small effects were observed in basic statistical models and not in adjusted statistical models. Thus, the small effects may be entirely explained by false-positives or confounding.

*4.2 Selective Reporting*

Researchers have flexibility to use different methods during a study. They also have flexibility to only report those methods that yield positive results and disregard those that yield unfavorable results (Kavvoura et al., 2007; Ioannidis, 2008; Contopoulos-Ioannidis et al., 2009; Ioannidis et al., 2011; Carp, 2012). This reporting preference involves a selective bias to highlight statistically significant findings and to avoid nonsignificant (null) findings in research (Kavvoura et al., 2007). What is reported can be challenging for public health policy makers to use because the significant findings may turn out to be nothing more than chance results (false positives).

An example of selective reporting (also called selective outcome reporting) is apparent with the Raju et al. (2000) review. They refer to a Paulin et al. (2017) randomized control trial of 30 children aged 5−12 years and state: "*Paulin and colleagues demonstrated that daily changes in household $NO_2$ exposure were associated with gas stove/oven use and led to worsened asthma symptoms and nighttime inhaler use among children with asthma*". Yet Paulin et al. state in their abstract: "*There were no associations between $NO_2$ and lung function or asthma symptoms*".

Paulin et al. used bivariate statistical models and multivariate statistical models (adjusted for gender, age, caregiver education, season). They examined four different outcomes between measured 24-hour indoor $NO_2$ level and possible asthma symptoms in their data set. It is noted that moving adjustment factors (covariates) into and out of statistical analysis models – i.e., selectively controlling for covariates – is a form of p-hacking (Simmons et al., 2011; John et al., 2012; Andrade, 2021). This relaxed practice can alter statistical significance between cause ($NO_2$) and effect (lung function or asthma symptoms).

Only one of the four Paulin et al. outcomes was reported by Raju et al. (2000) – nighttime inhaler use frequency – and it was statistically significant and was related to $NO_2$ levels in both statistical models. Whereas the other three outcomes that Paulin et al. examined – occurrence of daytime asthma symptom determined by twice-daily measured $FEV_1$, daytime inhaler use frequency, and nighttime awakening frequency – were not. It is unknown why Raju et al. did not mention these null findings.

Another example of selective reporting is with both Raju et al. and Garcia et al. reviews. Both repeatedly discussed various child asthmatic studies in their reviews. However, both failed to acknowledge the Wong et al. (2013) ISAAC cooking fuel−asthma study and its null findings. Wong and other researchers across the world investigated the association between asthma and use of various cooking fuels, including gas stoves, as part of the International Study of Asthma and Allergies in Childhood (ISAAC). ISAAC researchers collected data on 512,707 primary and secondary school children from 108 cities in 47 countries between 1999 and 2004.

Wong and other ISAAC researchers specifically examined and presented results for two statistical analysis models (initial adjusted models, final multivariate models) for gas stove−childhood asthma associations. Also, they examined two asthma outcomes (current symptoms of severe asthma and had asthma ever) for two age groups (6–7-year-olds and 13–14 year-olds) in these statistical models. All model outcomes were non-significant. Wong et al. stated in their abstract: "*we detected no evidence of an association between the use of gas as a cooking fuel and either asthma symptoms or asthma diagnosis*".

A third example is with the Lin et al. (2013a) meta-analysis itself. The Wong et al. (2013) ISAAC study was published 31 May 2013, almost three months prior to the Lin et al. meta-analysis (published 20 August 2013). The second listed co-author of both publications was the same person – B. Brunekreef, a senior researcher from Utrecht University, Utrecht, The Netherlands. It is unknown why Lin et al. failed to acknowledge the 2013 Wong et al. ISAAC study and its null findings, which Brunekreef (as a co-author of both studies) was aware of.

*4.3 P-hacking in Statistics*

P-hacking is a form of multiple hypothesis testing involving the search for significance during statistical analysis of data (Simmons et al. 2011; Hubbard, 2015; Harris, 2017, Streiner, 2018; Barnett & Wren, 2019; Moss & De Bin, 2021). P-hacking allows researchers to find statistically significant results with their data set even when studying a non-existent effect or association (Simonsohn et al., 2014).

Examples of p-hacking strategies in research have been reported in literature. Some examples include (Hahn et al, 2000; Simmons et al. 2011; John et al., 2012; Motulsky, 2015; Teixeira, 2018; Andrade, 2021; Stefan & Schonbrodt, 2023):

- Selectively choosing among dependent variables.
- Selectively manipulating/transforming predictor or outcome variables.
- Choosing sample size.
- Data trimming (selective exclusion of data outlying points).
- Using many covariates or selectively controlling for covariates.
- Selectively comparing different outcome variables.
- Using a variety of statistical analysis models.
- Selective reporting of significant results.
- Transforming a variable, for example by computing its logarithm or reciprocal.
- Conducting re-analysis on subsets of a data set.

The last procedure, conducting re-analysis on subsets of a data set (Hahn et al, 2000; Andrade, 2021; Stefan & Schonbrodt, 2023) was a feature of the Lin et al (2013a) meta-analysis – see Section 2.1. When a data set is subdivided into subgroups and if a particular subgroup is re-analyzed, it can easily find via diligent searching statistically significant results by chance. Pocock et al. (2004) note that in epidemiology investigations "*there is an increased risk of false claims of effect modification when several subgroup analyses are explored*" … and … "*there *is a need to exercise restraint, viewing subgroup findings as exploratory and hypothesis generating rather than definitive*".

Re-analysis on subsets of a data set by Lin et al. is also consistent with fishing expeditions – testing associations between combinations of variables in a data set with the hope of finding (and subsequently reporting) something statistically significant (Cormier & Pagano, 1999; Streiner, 2015; Andrade, 2021). Fishing expeditions tend to ignore some basic toxicological concepts like biological plausibility, magnitude of exposure to an investigated variable (e.g., gas stoves), genetics, locations of subjects, proximity to other causes, etc. When exposures are poorly defined and alternative explanations for a statistically significant association are ignored in these exercises, such associations would only be considered hypothesis generating and not proof of causation.

Another problem with meta-analysis is that researchers are further away from data collection than are the original researchers of the base papers. This separation, combined with multiple base papers used for meta-analysis, makes it impracticable for meta-analysts, peer reviewers and readers of a meta-analysis to understand data quality and method limitations of the base papers unless they evaluate the base papers themselves (Nelson et al., 2018). Nelson et al. state that: "*meta-analytic thinking not only fails to solve the problems of p-hacking, reporting errors, and fraud, it dramatically exacerbates them*".

Part of the problem is that base paper researchers have flexibility to only report methods that yield favorable results and to disregard those that yield unfavorable results (see Section 4.2). Recall that a hypothesis of meta-analysis is that test statistics taken from base papers are unbiased estimates of cause−effect (Boos & Stefanski, 2013). This is clearly not the case in practice given researcher flexibility to ignore reporting unfavorable results.

Another part of the problem is that when multiple testing is at play in the base papers, combining test statistics from these papers in meta-analysis can further worsen the effects of p-hacking. Multiple testing was a common feature of base papers used in the Lin et al. meta-analysis (see Table 1). On this problem, Nelson et al. state that: "*The end result of a meta-analysis is as strong as the weakest link; if there is some garbage in, then there is only garbage out*".

Multiple testing was a common feature of base papers used by Lin et al. in their meta-analysis. Re-analysis on subsets of a data set was a feature of the Lin et al meta-analysis itself. Given these features, multiple testing and p-hacking cannot be ruled out for test statistics presented by Lin et al. in support of their gas stove cooking−current asthma association claim.

*4.4 Implications*

The American Thoracic Society (Thurston et al., 2020) acknowledges an unclear (unproven) role of $NO_2$ in explaining a causal link with childhood asthma. The Lin et al. (2013a) meta-analysis used test statistics from epidemiology cause−effect studies (base papers) founded on a concept that indoor $NO_2$ may be bad and could cause childhood asthma, not on a concept of biological plausibility.

Our evaluation revealed methodological biases in the base papers used by Lin et al. and in the meta-analysis itself. Base papers used by Lin et al. show evidence of multiple testing (large numbers of hypothesis tests conducted). None of the 14 base papers we reviewed made any mention of this bias nor did they provide any explanation if no multiple testing procedure was used. The relaxed practice of multiple testing makes these types of studies likely to discover significant but false-positive associations. As to the meta-analysis itself, its methods show evidence of p-hacking.

P-value plots for both gas stove–childhood asthma and gas stove–childhood wheeze effects – Figures 2 and 3– show random (chance) associations, not real effects. The Lin et al. meta-analysis finding of a 'significant' gas stove–current asthma effect was not reproduced in a p-value plot (Figure 2). Whereas the meta-analysis finding of a nonsignificant gas stove–current wheeze effect was reproduced in a p-value plot (Figure 3).

Published literature reviews and observational studies discussed here highlight inconsistent findings and selective reporting of gas stove–childhood respiratory health associations. Inconsistent findings – i.e., both positive and negative (null) associations – may, in part, be due to a weak or indeed non-existent relationship between gas stove cooking (including $NO_2$) and childhood asthma outcomes.

Inconsistent findings may also be partly due to flexibility (biases) in researcher methods – i.e., multiple testing, p-hacking, and selective reporting (i.e., disregarding of results of studies with null findings). Goodman et al. (2016) note that multiple testing (multiplicity) and incomplete (selective) reporting might be the largest contributor to the occurrence of nonreproducibility (falseness) of published research claims.

Epidemiologists are not immune to unethical behavior (Bonita et al., 2006). Insufficient attention is paid by them to flexibility (biases) in their methods. This along with the practice of searching out and reporting weak statistical associations increases the potential for distorting the influences of chance and confounding in studies they publish (Bofetta et al., 2008).

Biases in epidemiology methods discussed here can be remedied with honest efforts to improve research transparency. With regard to gas stove–childhood respiratory health studies, this includes providing transparent details about procedures and data, statistical or analytical methods to test hypotheses, and completeness of reporting.

*Concluding remarks* – The Lin et al. (2013a) meta-analysis fails to provide reliable evidence for public health policy making on gas stove harms to children in North America. $NO_2$ is not established as a biologically plausible explanation of a causal link with childhood asthma. Biases – multiple testing and p-hacking – cannot be ruled out as explanation for a gas stove−current asthma association claim. A p-value plot for gas stove–childhood asthma shows a random (chance) association, not a real effect. The Gruenwald et al. (2023) estimate of current childhood asthma PAF due to gas stove cooking in the US is without substantiation and should be disregarded.

**Acknowledgments**

This research received no external funding.

**ORCID iD**

Warren B. Kindzierski: https://orcid.org/0000-0002-3711-009X

S. Stanley Young: https://orcid.org/0000-0001-9449-5478